\documentclass[pra,aps,twocolumn,showpacs,preprintnumbers,floatfix,10pt]{revtex4-1}
\usepackage{graphicx}
\usepackage{dcolumn}
\usepackage{bm}
\usepackage{latexsym,epsfig}
\usepackage{graphicx}
\usepackage{verbatim}
\usepackage{comment}
\usepackage{amsmath}
\usepackage{amssymb}
\usepackage{stmaryrd}
\usepackage{color}

\usepackage[colorlinks=true,linkcolor=blue,citecolor=blue]{hyperref}

\begin{document}
\title{Enhanced visibility of the Fulde-Ferrell-Larkin-Ovchinnikov state in one dimensional Bose-Fermi mixtures near the immiscibility point}
\author{Manpreet Singh}
\affiliation{Universit\'{e} de Paris, Laboratoire Mat\'{e}riaux et Ph\'{e}nom\`{e}nes Quantiques, CNRS, F-75013, Paris, France}
\author{Giuliano Orso}
\email{giuliano.orso@univ-paris-diderot.fr}
\affiliation{Universit\'{e} de Paris, Laboratoire Mat\'{e}riaux et Ph\'{e}nom\`{e}nes Quantiques, CNRS, F-75013, Paris, France}

\date{\today}

\begin{abstract}
Based on the matrix product states method, we investigate numerically the ground state properties of one-dimensional 
mixtures of repulsive bosons and spin-imbalanced attractive fermions, the latter being in the 
Fulde-Ferrell-Larkin-Ovchinnikov (FFLO) state, where Cooper pairs condense at a finite momentum $k=k_{FFLO}$.
We find that the visibility of such a state is dramatically enhanced as the repulsive Bose-Fermi mixture is brought close 
to the phase-separation point. In particular, large amplitude self-induced oscillations with 
wave-vector $2k_{FFLO}$ appear in both the  fermion total density and the boson density profiles, leaving sharp
fingerprints in the corresponding static structure factors. We show that these features remain well visible in cold atoms
systems trapped longitudinally by a smooth flat-bottom potential. 
Hence bosons can be used to directly reveal the modulated  Fermi superfluid in experiments.
\end{abstract}

\maketitle

\section{Introduction}
According to the Bardeen-Cooper-Schrieffer (BCS) theory of superconductivity, electrons with opposite spin bind into
bosonic pairs, which then condense in the state of zero center-of-mass momentum, leading to macroscopic phase coherence
and vanishing electrical resistance. An intriguing question is the possible coexistence of superconductivity with a 
spin imbalance, which destabilizes the BCS mechanism. Several exotic superfluid states have been proposed theoretically,
including the breached pair or Sarma state \cite{sarma19631029,vincentliu1,vincentliu2}, states with deformed Fermi
surfaces \cite{sedrakian1,sedrakian2} and Fulde-Ferrell-Larkin-Ovchinnikov (FFLO) state \cite{fulde64, larkin64}, to name
a few.

The FFLO state is characterized by the condensation of Cooper pairs at finite momentum $\textbf k_{FFLO}$, corresponding
in real space to a spatially modulated order parameter. As a consequence, excess fermions, which are detrimental to
superconductivity, are stored preferentially at the nodes of the pairing field, leading to an oscillation in the spin
density with wave-vector $2 \textbf k_{FFLO}$ (corresponding to two nodes per wavelength). The FFLO state is currently
being investigated in a variety of physical systems, including layered organic~\cite{KoutroulakisPRL2016}, heavy fermion
~\cite{PfleidererRMP2009} and iron-based~\cite{PtokJLTM2013} superconductors, hybrid superconducting-ferromagnetic
structures~\cite{buzdinRMP2005} and quark-gluon plasma~\cite{nardulli04}. To date, its experimental evidence relies
mostly on thermodynamic measurements, although recent NMR spectra of organic superconductors are consistent with a
periodic modulation of the spin density~\cite{KoutroulakisPRL2016}.

Spin-imbalanced atomic Fermi gases~\cite{Zwierlein2006,Partridge2006} provide an alternative route to investigate the
FFLO state, especially in one dimensional (1D) geometries, where the exact Bethe ansatz solution of the
microscopic model allow to derive~\cite{orso1,drummond} the grand canonical phase
diagrams for both homogeneous and trapped systems.  
For a nonzero attractive contact interaction, the ground state of the spin-imbalanced 
system is FFLO-like, as confirmed by several numerical~\cite{meisner,fazio,batrouni,ueda,mueller1,meisner2010} and
analytical~\cite{GuanPRB2007,bolech,LeeNPB2011,Cheng:PRB2018} studies (for a review see~\cite{reviewZwerger:2011,chaohong}). 
1D systems with combined spin and mass 
imbalances have also been shown to exhibit  the modulated superfluid phase~\cite{Wang:PRA2009,Batrouni_2009,Burovski:PRL2009,Orso:PRL2010,Roux:PRA2011,Dalmonte:PRA2012,PhysRevA.101.033603,rammelmller2020pairing}. 

While the predicted two-shell structure of the density profiles  
was soon confirmed~\cite{mueller2} experimentally,  no evidence of the periodic modulation of the spin density was found. 
Several detection schemes  have been put forward since then (for a review see~\cite{torma1}), which are based on the analysis of collective oscillations
~\cite{cooper}, the sudden expansion of the gas~\cite{KajalaPRA2011,LuPRL2012,BolechPRL2012}, interaction quenches
~\cite{orso2}, noise correlations~\cite{lauchli,pecak2019signatures}, spectroscopy measurements~\cite{torma3,torma4,roscilde,LutchynPRA2011} 
and interference techniques~\cite{trivedi}.

Atomic Bose-Fermi mixtures provide a natural playground for several quantum phenomena~\cite{anders2012}, including double
superfluidity~\cite{ferrier2014,delehaye2015,Yao2016,AdhikariPRA2010,danshita2013,OzawaPRA2014,Wen_2018}, phase-separated
states and interfaces~\cite{Lous2018,DehkharghaniNJP2017,SieglPRA2018,Decamp2016}, supersolidity
~\cite{BuchlerPRL2003,HebertPRB2008,TitvinidzePRB2009}, pairing from induced interactions
~\cite{HeiselbergPRL2000,matera2003,demler,albus,zhai,mathey2007,enss2009,Pasek2019} or in mixed dimensions
~\cite{Wu2016,OkamotoPRA2017,Caracanhas2017,KinnunenPRL2018}. 
Recently, Ref.~\cite{MelkaerPRA2018} investigated a two-dimensional spin-imbalanced Fermi gas immersed in a Bose superfluid, 
leading to an effective long-range attractive interaction between fermions. The authors showed that the  FFLO state 
can occupy a larger portion of the ground state phase diagram as compared to the case of fermions with direct contact interactions. 
In 1D systems,  however, the FFLO state is energetically stable already in the absence of bosons.

In this work we suggest that bosons can instead be used as a sensitive probe of the exotic superfluid, 
once the repulsive Bose-Fermi mixture is brought sufficiently close to the phase-separation point.
Specifically, we find that robust self-induced density modulations with wave-vector $2 k_{FFLO}$  suddenly appear both in
the boson and in the fermion total density profiles, leading to sharp kinks in the corresponding static structure factors
(much sharper than the original kink in the magnetic response). 
The new phenomenon is completely general and can be observed experimentally with ultracold atoms  confined in smooth flat-bottom traps.

The paper is organized as follows. In Section~\ref{sec:model} we introduce the microscopic model for the Bose-Fermi mixture and the numerical method used to study it. In Section~\ref{sec:stability} we verify that the ground state of the spin imbalanced Fermi gas remains FFLO-like, even in the presence of bosons.
Section~\ref{sec:visibility} describes the main result of our paper, namely
the boson-induced enhancement of FFLO visibility near phase separation. In
Section~\ref{sec:trap} we proove that the observed phenomenon persists also when the mixture is confined in a smooth flat-bottom trap.  Section~\ref{sec:negativeUbf} shows that for attractive Bose-Fermi interaction the effect is barely visible, pointing out the limitations of previous perturbative models for the mixture. Finally, Section~\ref{sec:conclusion} provides  a summary and an outlook.

\section{Model and method}
\label{sec:model}
We describe a homogeneous Bose-Fermi mixture by the following lattice Hamiltonian:
\begin{eqnarray}
H=&-&t_f\sum_{<i,j>,\sigma}c_{i\sigma}^{\dagger} c_{j\sigma}+U_{f}\sum_{i}n_{i\uparrow}n_{i\downarrow}-t_b\sum_{<i,j>}b_{i}^{\dagger} b_{j} 
  \nonumber\\
&+&\frac{U_{b}}{2}\sum_{i}n_{ib}(n_{ib}-1)+ U_{bf}\sum_{i}n_{ib}(n_{i\uparrow}+n_{i\downarrow}).
\label{eq:one}
\end{eqnarray}
where the  first two terms represent the Fermi Hubbard model, $c_{i\sigma}^{\dagger} $ being the local
creation operators for fermions with spin component ${\sigma=\uparrow,\downarrow}$,  $t_f$ is their tunneling
rate  and $U_f$ $(U_f<0)$ is the strength of the attractive interaction between fermions with opposite spin. The third and fourth terms yield the Bose-Hubbard model, where ${b^\dagger_i}$ is
the bosonic creation operator at site $i$, while $t_b$ and $U_b$ $(U_b>0)$ are the corresponding tunneling rate
and the onsite repulsion strength. Bosons and fermions are coupled by repulsive contact interactions of strength $U_{bf}$
$(U_{bf}>0)$, as described by the last term in the rhs of Eq.(\ref{eq:one}). In the following we assume
that bosons and fermions have equal tunneling rates and fix the energy scale by setting $t_f=t_b=1$.

Our numerical results are based on the Density Matrix Renormalization Group (DMRG) method, expressed in terms of Matrix
Product States (MPS) (for a review see Ref.~\cite{schollwoeck2011}). Specifically, we use the \emph{mps-optim} code of
the ALPS library~\cite{dolfi2014}. We consider a chain of $L=120$ sites with open boundary conditions, containing
$N_{\uparrow}=40$ spin-up fermions, $N_{\downarrow}=32$ spin-down fermions and $N_{b}=60$ bosonic atoms (the choice of
half filling for bosons is not crucial for our results). We set $U_{b}=4$ and $U_{f}=-2$ while varying the Bose-Fermi
coupling $U_{bf}$. To ensure proper convergence, we allow a maximum occupancy of four bosons per-site along with bond
dimension up to $4000$ and $80$ sweeps.

\begin{figure}
	\centering
	\includegraphics[width=0.92\columnwidth]{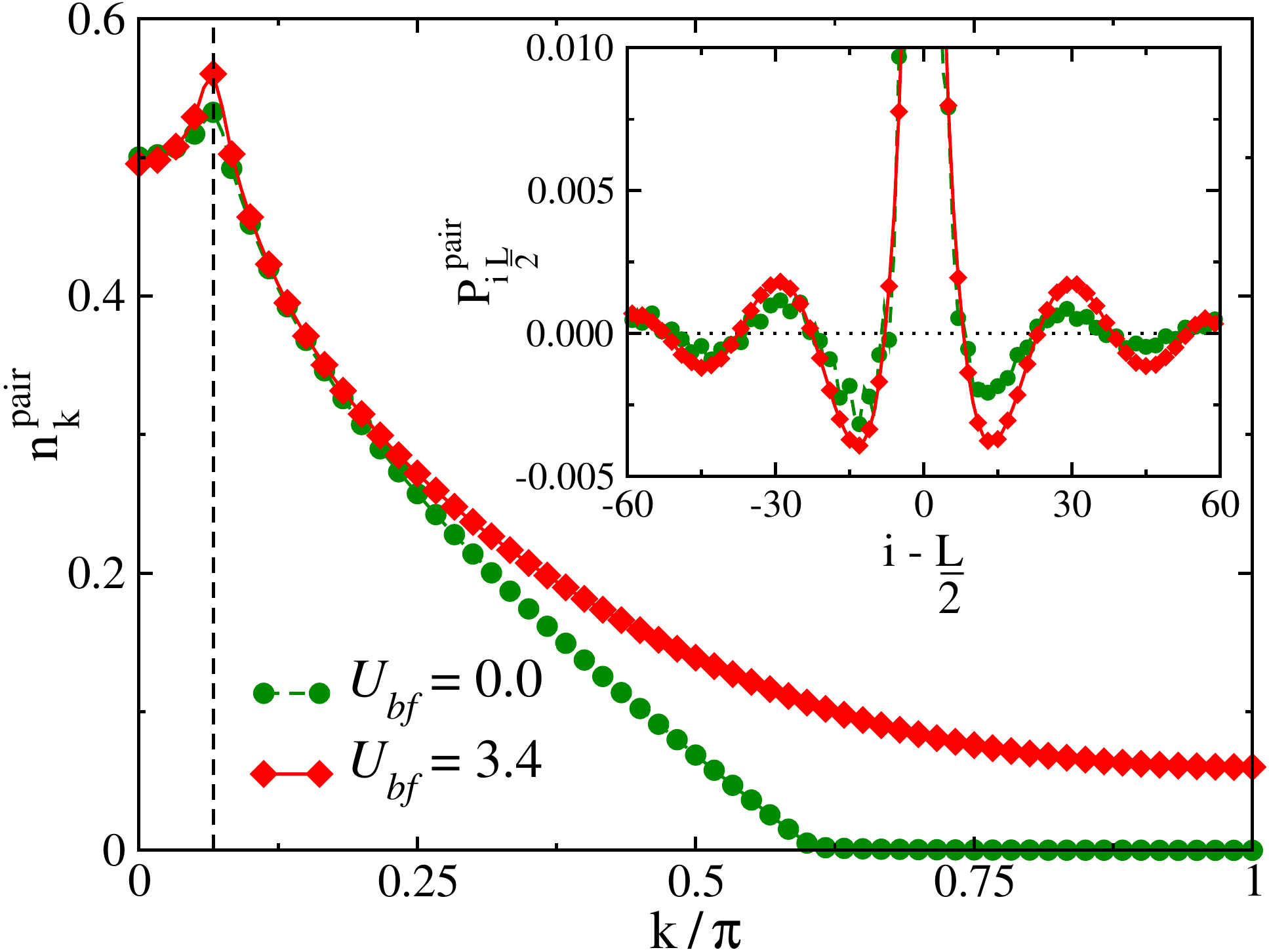}
	\caption{(Color online) Pair momentum distribution, cf.~Eq.(\ref{eq:three}), for $U_{bf}=0$ (green circles) and for
		$U_{bf}= 3.4$ (red diamonds), where the Bose-Fermi mixture is close to phase separation. The dashed line marks the
		position of the FFLO wave-vector $k_{FFLO}=k_{F\uparrow}-k_{F\downarrow}=\pi/15$. The inset shows the corresponding
		results for the singlet superconducting correlation function $P_{ij}^{pair}$, cf.~Eq.(\ref{eq:two}), with $j=L/2$ (center
		of the chain), as a function of the distance $i-L/2$ (for clarity odd sites data have been skipped). The dotted line
		marks the zero crossing. The intraspecies interaction strengths are $U_b=4$ and $U_f=-2$. The chain has size $L=120$ and
		contains $N_{\uparrow}=40$ spin-up fermions, $N_{\downarrow}=32$ spin-down fermions and $N_b=60$ bosons.}
	\label{fig:fig1_ppc_nkpair}
\end{figure}

\begin{figure*}
	\noindent
	\hspace{-0.7cm}
	\centering \includegraphics[width=19.5cm]{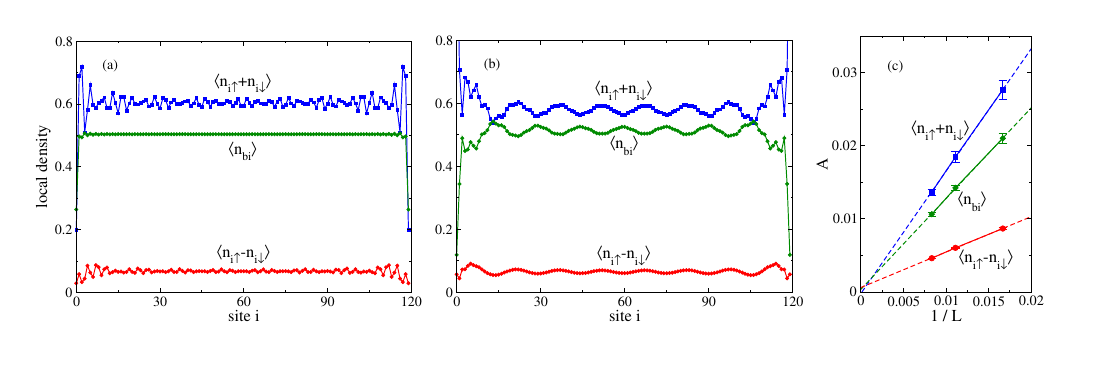}
	\vspace{-1.2cm}
	\caption{(Color online) Density profiles of the homogeneous mixture for $U_{bf}=0$ (panel a) and $U_{bf}=3.4$ (panel b).
		The three data curves correspond to the spin density (red circles), total fermionic density (blue squares) and boson
		density (green diamonds), respectively. The particle numbers are the same as in Fig.\ref{fig:fig1_ppc_nkpair}. The panel
		c shows the amplitudes of the FFLO oscillations in the three density profiles as a function of $1/L$ for $U_{bf}=3.4$ and
		for three different values of the system size, $L=60, 90$ and $120$ at fixed particle densities. The intraspecies
		interaction strengths are $U_f=-2$ and $U_b=4$.}
	\label{fig:fig2_denprof}
	\vspace{-0.3cm}
\end{figure*}

\section{Stability of the FFLO state}
\label{sec:stability}
We first show that the nature of the ground state of the spin-imbalanced Fermi gas remains FFLO-like even in the presence of 
bosons, as long as the homogeneous mixture remains stable. To this purpose, we define the pair momentum distribution (PMD) as
\begin{equation}
n_k^{pair}=\frac{1}{L} \sum_{i,j} e^{\mathrm{i}(i-j)k} P_{ij}^{pair},
\label{eq:three}
\end{equation}
where
\begin{equation}
P_{ij}^{pair}=\langle c_{i\uparrow}^{\dagger} c_{i\downarrow}^{\dagger} c_{j\downarrow} c_{j\uparrow} \rangle
\label{eq:two}
\end{equation}
is the superconducting correlation function in the singlet channel. Figure \ref{fig:fig1_ppc_nkpair} shows the PMD for
$U_{bf}=0$ (green circles) and for $U_{bf}=3.4$ (red diamonds). The
1D FFLO state is signaled by a sharp peak in the PMD at a finite momentum $k_{FFLO}=k_{F\uparrow}-k_{F\downarrow}$
(dashed line), where $k_{F\uparrow}=\pi N_\uparrow/L$ and $k_{F\downarrow}=\pi N_\downarrow/L$ are the Fermi momenta of
the majority and minority spin components.
 
We see from Fig.\ref{fig:fig1_ppc_nkpair} that the characteristic FFLO peak remains well visible in the presence of bosons and is even slightly taller, confirming the energetic stability of the FFLO phase. The main effect of the repulsive boson-fermion interaction is the appearance of Cooper pairs with
large momentum, close to the edge of the Brillouin zone. Since  $\sum_k  n_k^{pair}=\sum_i \langle n_{i\uparrow} 
n_{i\downarrow}\rangle $, this results in an increase of the number of doubly occupied sites from $17.4$ to $23.3$.
We have verified numerically that this behavior is not intrinsic to the FFLO state, as it also occurs for equal spin
populations, $N_\uparrow=N_\downarrow$. The PMD could be measured by projecting the Cooper pairs into deep molecular
states and performing time-of-flight experiments~\cite{yang2}, as previously done for three-dimensional Fermi 
superfluids, although interactions effects during the expansion can complicate the picture.

In coordinates space, the FFLO state appears as a self-generated spatial modulation of the superconducting correlation
function, Eq.(\ref{eq:two}), superimposed to the  algebraic decay typical of 1D systems. Our numerical results for
$P_{ij}^{pair}$ are displayed in the inset of Fig.\ref{fig:fig1_ppc_nkpair} for the aforementioned values of $U_{bf}$.
Interestingly, as $U_{bf}$ increases, we see that this quantity smoothens out and becomes more symmetric with respect to
the center of the chain, which might favors the observation of the exotic superfluid with interferometric techniques.

\section{Manifestation of FFLO order near phase separation} 
\label{sec:visibility}

\subsection{Density profiles}

Let us now investigate the much more interesting effects of the Bose-Fermi repulsion on the density profiles of the two species. With the spectacular recent
advances in quantum gas microscopy~\cite{Haller2015,Parsons2015,Cheuk2015}, it is now possible to measure these local
observables  with high resolution. In Fig.\ref{fig:fig2_denprof} we plot the distributions of the spin density $\langle
n_{i\uparrow}-n_{i\downarrow}\rangle$ (red circles), the total fermionic density 
$\langle n_{i\uparrow}+n_{i\downarrow}\rangle$ (blue squares) and the boson density $n_{i b}$ (green diamond) in the
absence of coupling (panel a) and close to phase separation (panel b). Although in 1D systems true long range order is
absent due to the strong  quantum fluctuations, the crystalline structure in the local spin density can appear in chains
of finite size.  

We see in Fig.\ref{fig:fig2_denprof} (a) that for $U_{bf}=0$ the FFLO oscillation is barely visible, as the attraction
between fermions is relatively weak, $U_f=-2$. Indeed, previous DMRG~\cite{fazio} and quantum Monte Carlo~\cite{roscilde}
studies have observed a clear periodic modulation only for relatively strong attractions between fermions, 
say $|U_f|\gtrsim 5$. A Fourier analysis of the data, presented in Appendix A, shows that the fermion total density displays
oscillations with wave-vectors $2k_{F\downarrow}$ and $2k_{F\uparrow}$, while the $2k_{FFLO}$ modulation is nearly
absent. In contrast, the boson density displays oscillations with dominant wave-vector $k=\pi$.

The situation for $U_{bf}=3.4$  is completely different. As displayed in Fig.\ref{fig:fig2_denprof} (b), 
the spin-density profile exhibits a clear periodic structure with the expected wavelength, $2\pi/(2 k_{FFLO})=15$.
Surprisingly, the same crystal order is  imprinted in the total fermionic density and in the  boson density profiles, while 
 the usual density modes with wave-vector $2k_{F\sigma}$ almost disappear, as discussed in Appendix A. This observation is the key result of our paper. 

We also see in Fig.\ref{fig:fig2_denprof} (b) that  the oscillation in the boson density is out of phase by a factor of $\pi$
with respect to the other two densities, due to the repulsive Bose-Fermi interaction.
Moreover at the edge of the chain there are more fermions than bosons, so in the bulk  the total fermionic  (bosonic) density oscillates around a lower (higher) mean value.

Since we expect that the density modulations displayed in Fig.\ref{fig:fig2_denprof} (b) are a \emph{direct manifestation} of the FFLO pairing, 
 their amplitudes $A$ must vanish in the limit of infinite chains. In order to verify this crucial 
 point, we have  performed similar calculations for
system size $L=60$ and $L=90$, keeping the  densities $N_\sigma/L$ and $N_b/L$ unchanged. We compute $A$ by fitting the
numerical data in the central region of the chain, corresponding to $L/4$ sites, with a function $n(x)=A \cos(2k_{FFLO}x+\phi)$. The result
is shown in Fig.\ref{fig:fig2_denprof} (c) as a function of $1/L$. All the data curves are well fitted by straight lines
with approximately zero intercept, thus confirming our claim.

\begin{figure}
\centering
\includegraphics[width=0.4\textwidth]{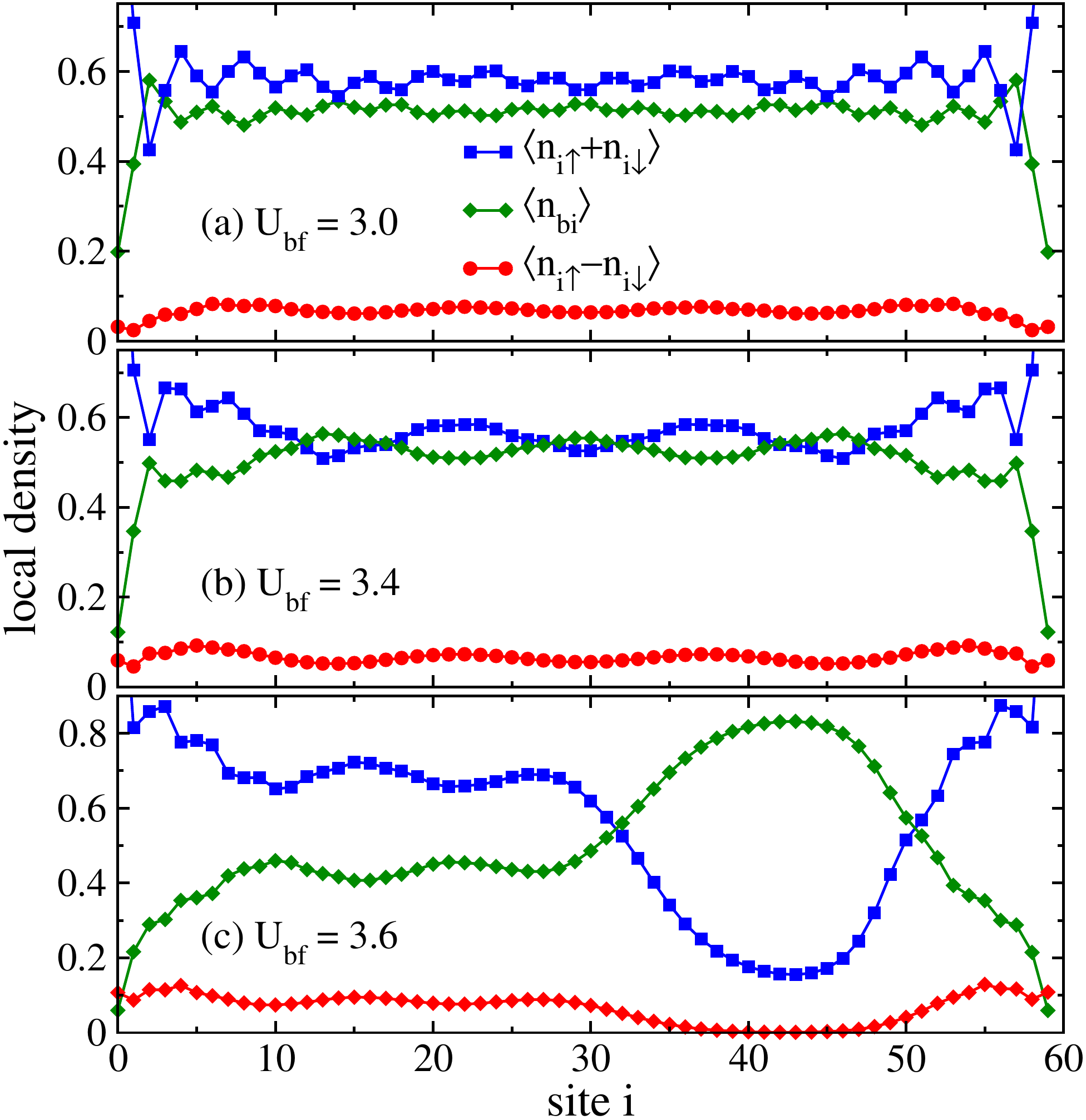}
\caption{(Color online) Local density profiles of bosons and fermions calculated for $U_{bf}=3$ (panel a), 
$3.4$ (panel b) and $3.6$ (panel c). The intraspecies interaction strengths are set to $U_{f}=-2$ and $U_{b}=4$ (same as
in the main text). The system size is $L=60$ and the particle numbers are re-scaled to $N_\uparrow=20, N_\downarrow=16$
and $N_b=30$ in order to keep the overall densities constant. The three data curves correspond to the spin density (red
circles), total fermion density (blue squares) and boson density (green diamonds), respectively.}
\label{fig:phase_sep1}
\end{figure}

Let us now clarify under which conditions the FFLO nodal structure is imprinted on the density distributions of the two species.
 While the FFLO modulation in the spin density profile appears progressively as the boson-fermion repulsion becomes stronger,	the corresponding effect in the total fermion density and in the boson density profiles appear \emph{only}  when
	the system is close enough to the phase separation point.	
This key fact is illustrated in  Fig.\ref{fig:phase_sep1}, where we display the density profiles of the mixture for increasing values of the Bose-Fermi repulsion strength $U_{bf}$ calculated for a chain of $L=60$ sites.
For $U_{bf}=3$ (panel a), the spin density displays FFLO oscillations, but  the other two density profiles show modulations with  shorter wave-lengths, corresponding to the usual $2k_{F\uparrow}, 2k_{F\downarrow}$ modes.  
For $U_{bf}=3.4$  (panel b) all three density profiles oscillate with the same $2k_{FFLO}$  wave-vector, as also displayed in Fig.\ref{fig:fig2_denprof}.
Further increasing the boson-fermion repulsion leads to an instability of the homogeneous mixture towards phase separation, 
as shown in  Fig.\ref{fig:phase_sep1} (c) for $U_{bf}=3.6$.

We have verified numerically that the FFLO imprinting observed in Fig.\ref{fig:fig2_denprof} (b) is a completely general phenomenon, which occurs also for weak boson-boson repulsion or for strong fermion-fermion attraction, as long as the system is close to the immiscibility point. More details can be found in Appendix B.

\subsection{Static structure factors}
\begin{figure}
\centering
\vspace{-0.1cm}
\includegraphics[width=0.95\columnwidth]{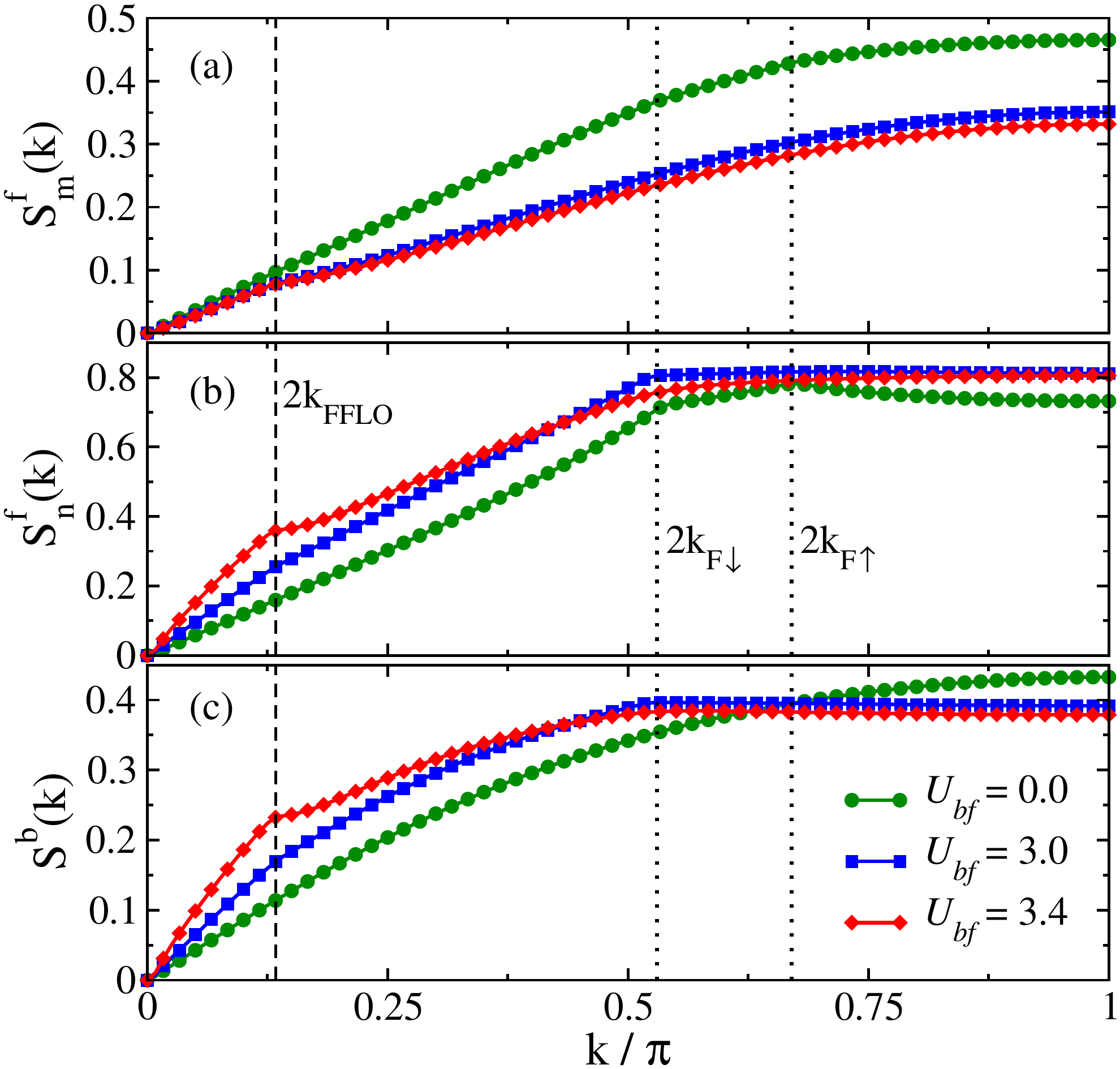}
\vspace{-0.1cm}
\caption{Three different types of static structure factors plotted as a function of momentum:
(a) fermion spin density ($S_m^f$), (b) fermion total density ($S_{n}^f$), and (c) boson density ($S^b$). The three
curves in each panel correspond to $U_{bf}=0.0$ (green circles), $3.0$ (blue squares) and $3.4$ (red diamonds). The
dashed vertical lines mark $2k_{FFLO}$, while the two dotted lines correspond to $2k_{F\downarrow}$ and $2k_{F\uparrow}$
(see text for details).}
\vspace{-0.3cm}
\label{fig:fig3_ssf}
\end{figure}

The static structure factors of the density distributions of bosons ans fermions
can be accessed experimentally in cold atoms samples via Bragg scattering~\cite{HoinkaPRL2013,PhysRevLett.121.103001} or
quantum polarization spectroscopy~\cite{roscilde}. They are defined as
\begin{eqnarray}
S^b(k)&=& \sum_{i,j} e^{\mathrm{i}(i-j)k} (\langle n_{ib} n_{jb} \rangle - \langle n_{ib} \rangle \langle n_{jb} \rangle), \nonumber\\
S_{O}^f(k)&=& \sum_{i,j} e^{\mathrm{i}(i-j)k} (\langle O_i O_j \rangle - \langle O_i \rangle \langle O_j \rangle),
\label{eq:five}
\end{eqnarray}
where for fermions we distinguish between the spin response, corresponding to the operator 
$O=m=n_{\uparrow}-n_{\downarrow}$, and the total density response, where $O=n=n_{\uparrow}+n_{\downarrow}$.
Fig.\ref{fig:fig3_ssf} displays the momentum dependence of the three static structure factors for three different values
of the Bose-Fermi coupling, $U_{bf}=0$ (green circles), $3.0$ (blue squares) and $3.4$ (red diamonds).

For $U_{bf}=0$ we see that both $S_{m}^f$ and $S_{n}^f$  exhibit similar shapes with kinks at $k=2k_{F\downarrow}$ and
$k=2k_{F\uparrow}$ (dashed lines). This can be understood by noticing that in a  noninteracting Fermi gas the two static
structure factors coincide, $S_m=S_n$, and are given by~\cite{roscilde}
\begin{eqnarray}\label{eq:ssf}
S_m(k)&=&|k|/\pi\;\;\;\textrm{for}\;0<|k|<2k_{F\downarrow}\nonumber \\
&=&|k|/(2\pi)\;\;\;\textrm{for}\;2k_{F\downarrow}<|k|<2k_{F\uparrow} \nonumber \\
&=&(k_{F\uparrow}+k_{F\downarrow})/\pi\;\;\; \textrm{for}\; |k|>2k_{F\uparrow}.
\end{eqnarray}  
The inclusion of a moderate attraction between fermions slightly smears the two kinks and decreases (increases) the overall scale of the magnetic (total density) structure factor, as indicated by the green curves in the panels (a) and
(b). In contrast, the density response of the Bose gas is smooth and increases monotonously as the momentum increases, as
shown by the panel(c) of the same figure.

As $U_{bf}$ increases, we see that a kink progressively stands out in the spin response at $k=2k_{FFLO}$, signaling the
FFLO state. At the same time  both the total density and the boson density responses become approximately flat for
$k>2k_{F\downarrow}$. 
By further approaching the immiscibility point, these two quantities develop a sharp kink at $k=2k_{FFLO}$, as shown in
Fig.\ref{fig:fig3_ssf} (b) and (c). In particular, we see that the two kinks are significantly more pronounced than the
corresponding one in the original magnetic response, thus favoring the detection of the FFLO state. We emphasize that the
results shown in Fig.\ref{fig:fig3_ssf} have negligible finite size effects, as shown in Appendix A. This is due to the fact that
the static structure factors measure density correlations at different sites, which remain finite in the thermodynamic
limit.

\section{Effect of a smooth flat-bottom trapping potential}
\label{sec:trap}
So far we have considered homogeneous mixtures confined in a box with open boundary condition. The effects of a smooth trapping potential acting on bosons and fermions can be taken into account through the generalized
Hamiltonian
\begin{equation}
H_\textrm{trap}= H+ \sum_{i} V \left (i-\frac{L}{2}\right )^p(n_{ib}+n_{i\uparrow}+n_{i\downarrow}), 
\end{equation}
where $V$ and $p$ are positive numbers. Since the FFLO wave-vector is fixed by the value of the local spin density, the
latter should stay approximately constant over a wide region of the trap for the corresponding density modulations to be
observable. Hence a confinement sharper than harmonic, $p>2$, is generally required. Flat-bottom potentials for ultracold atoms can be
realized optically, by using a digital micro-mirror device (DMD); for instance $p\simeq 16$ in the experiment of Ref.~\cite{MukherjeePRL2017}.

\begin{figure}
\centering
\begin{tabular}{c}
\includegraphics[width=0.95\linewidth]{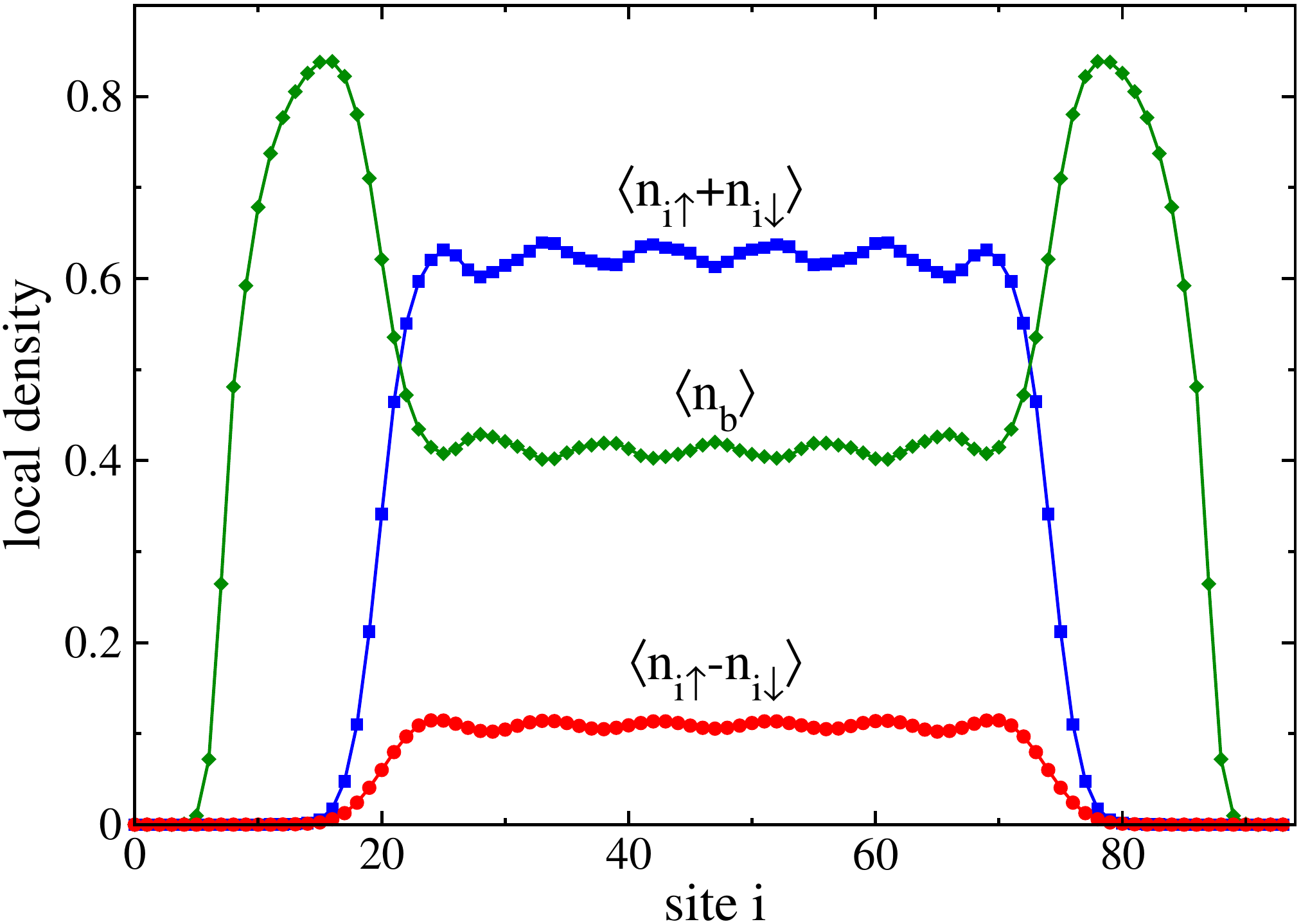}
\end{tabular}
\caption{(Color online) Spin density (red circles), total fermionic density (blue squares) and boson density (green
diamonds) profiles of atoms in a quasi flat trap with power $p=12$ and pre-factor $V=10^{-19}$. The interactions
parameters are $U_b=4$, $U_f=-2$ and $U_{bf}=3.4$. The mixture contains $N_\uparrow=20$ spin-up fermions, 
$N_\downarrow=14$ spin-down fermions and $N_b=42$ bosons.}
\label{fig:fig4_trap}
\end{figure}

In Fig.~\ref{fig:fig4_trap} we display the calculated density profiles for a mixture with 
$N_\uparrow=20, N_\downarrow=14$ and $N_b=42$ in a trap with $p=12$ and $L=94$, using the same values for the
interactions strengths as in Fig.\ref{fig:fig2_denprof} (b). We see that the characteristic FFLO modulations in the
density profiles of bosons and fermions are well visible in the middle of the trap and can therefore be used as evidence of a FFLO phase in this region.

\begin{figure}
\centering
\includegraphics[width=0.48\textwidth]{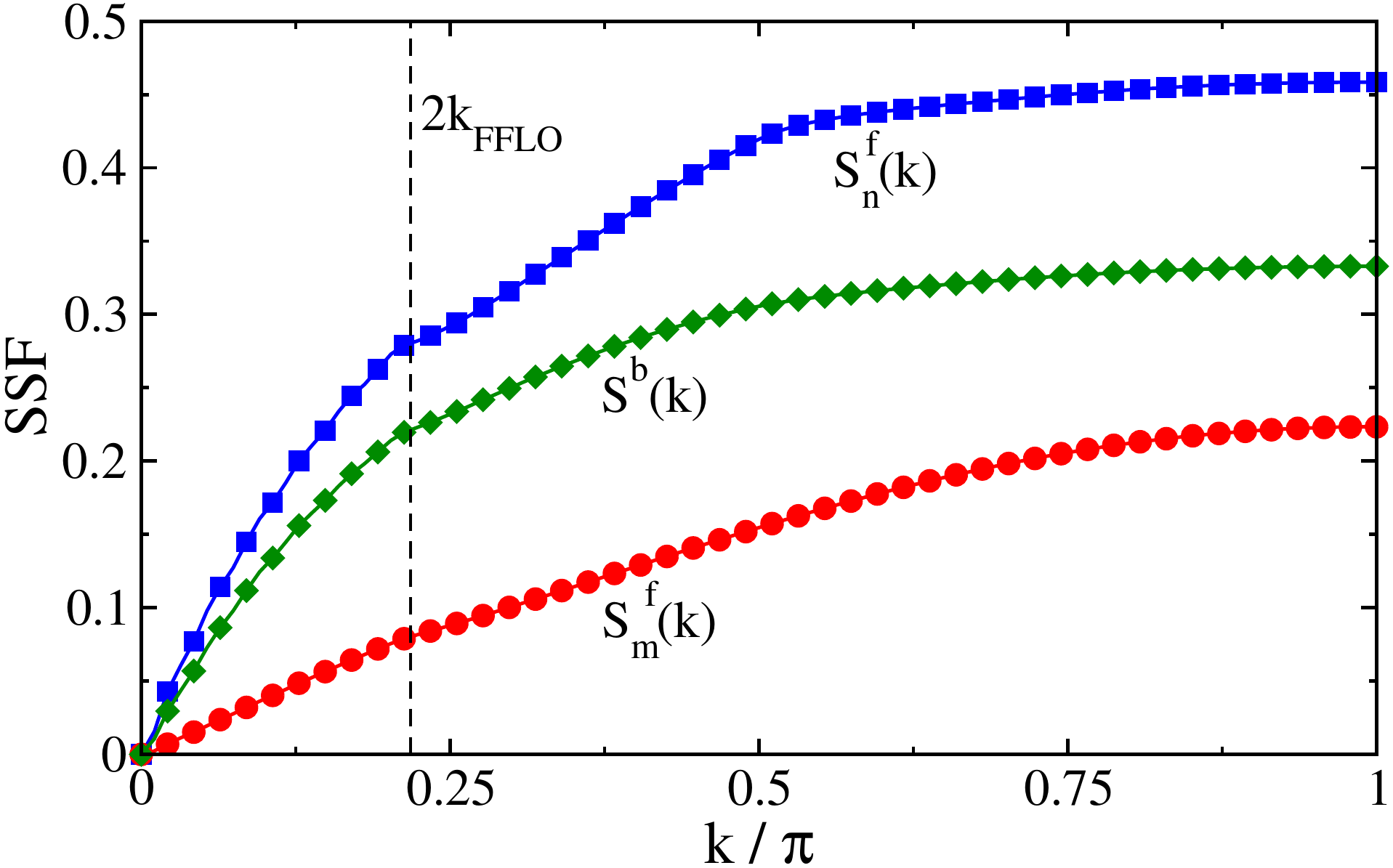}
\caption{(Color online)  Static structure factors  as a function of momentum calculated in the presence of the trap. All  
parameter values and particle numbers are as in Fig.\ref{fig:fig4_trap}.}
\label{fig:SSF_trap}
\end{figure}

In  Fig.\ref{fig:SSF_trap}
we display the corresponding results for the static structure factors in the presence of the smooth trap. We see that the FFLO kinks in $S^f_n$ and $S^b$ remain remarkably sharp, implying that the observed density modulations can be detected in cold atoms experiments.
On the other hand, the kink in the magnetic response is less evident, as we already observed for homogeneous mixtures.

 \section{Attractive Bose-Fermi mixture}
\label{sec:negativeUbf}
It is worth emphasizing that our mechanism of boson-enhanced FFLO visibility cannot be described using perturbative models of the mixture which are only valid for weak boson-fermion repulsion, far from the phase separation limit.
For instance in several mean field studies~\cite{HeiselbergPRL2000,matera2003,demler,albus,MelkaerPRA2018} of induced superfluid pairing in Bose-Fermi mixtures, 
bosonic degrees of freedom are integrated out using the adiabatic 
approximation of fast bosons. One is then left with a purely fermionic Hamiltonian, 
where particles are subject to an effective boson-induced long-range attractive interaction. 
Within second-order perturbation theory, the strength of this interaction  is proportional to $U_{bf}^2$, and is therefore insensitive to the sign of the boson-fermion coupling. Hence the approximate model predicts \emph{identical} effects for positive and negative values of $U_{bf}$.

\begin{figure}
\centering
\includegraphics[width=0.46\textwidth]{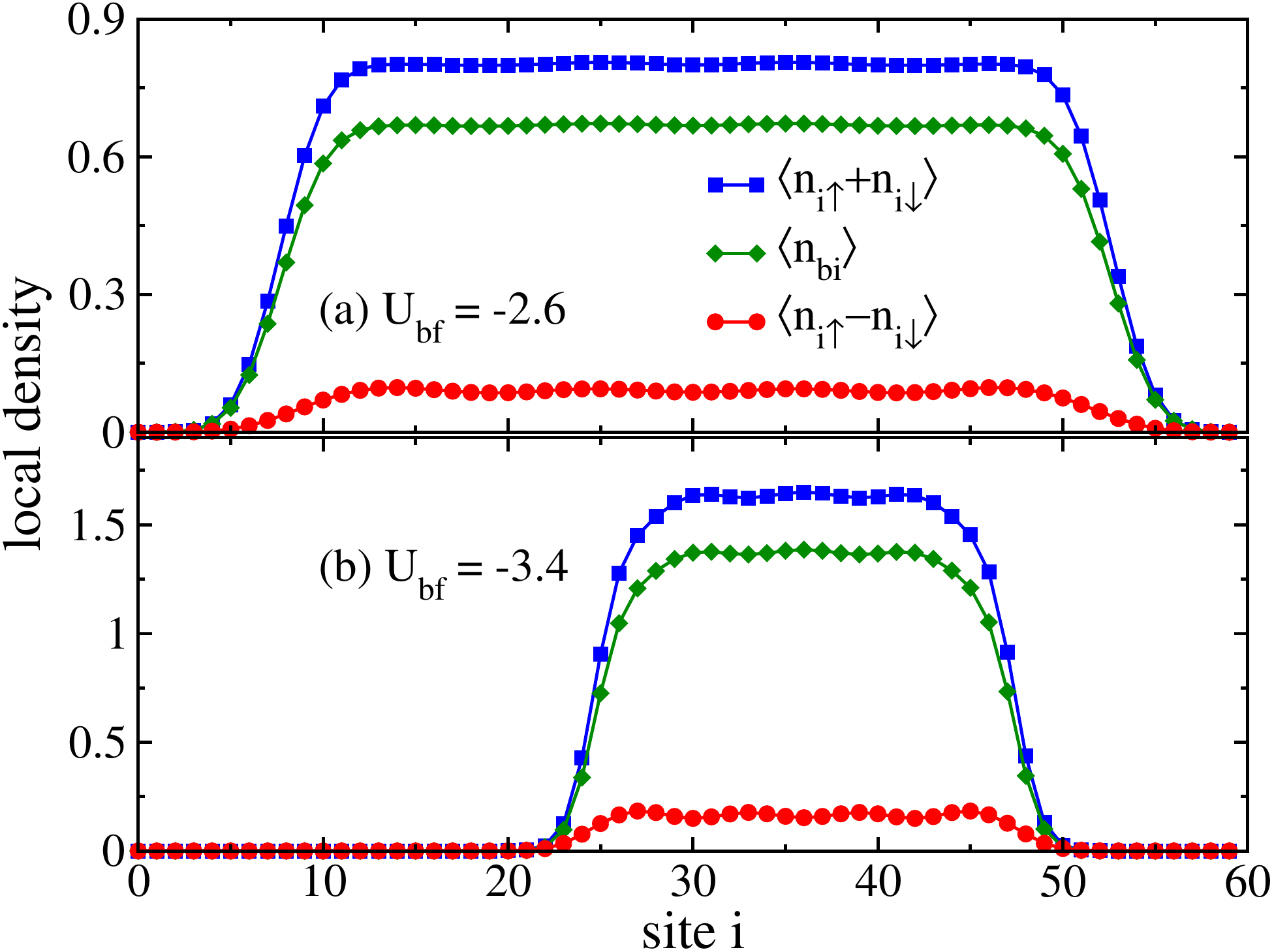}
\caption{(Color online) Local density profiles of bosons and fermions calculated for $U_{bf}=-2.6$ (panel a), and $-3.4$ (panel b). The values of the intra-species interaction strengths and the population numbers are the same as in Fig.\ref{fig:phase_sep1}. 
}
\label{fig:neg_Ubf}
\end{figure}

 This fact, however, is in stark contrast with our numerical findings for attractive Bose-Fermi mixtures, shown in Fig.\ref{fig:neg_Ubf}. The results for the density profiles are obtained using the same values of the intra-species interaction strengths and particle numbers as in Fig.\ref{fig:phase_sep1} but assuming $U_{bf}=-2.6$ (panel a) and $U_{bf}=-3.4$ (panel b).   
 We see that, already at $U_{bf}=-2.6$, the edges of the chain are no longer occupied, while for $U_{bf}=-3.4$ the densities distributions shrink to roughly half of the available lattices sites corresponding to a droplet-like phase.
 
 Interestingly, a zoom to the data of Fig.\ref{fig:neg_Ubf} reveals that, differently from the case of repulsive boson-fermion interactions, 
 the FFLO oscillations in the boson and total fermion densities are in phase, while the spin density is out of phase by a factor $\pi$.

\section{Conclusion and outlook}
\label{sec:conclusion}
In summary, we have investigated the ground state properties of a 1D FFLO state coupled to a Bose superfluid through strong repulsive interactions. 
While bosons do not affect the energetic stability of the FFLO phase (as long as the mixture remains homogeneous),
large amplitude modulations with wave-vector $2k_{FFLO}$ appear in  the  density profiles of the two species,
 as the mixture is  close enough to the immiscibility point. The same nodal structure is also imprinted in the density correlations,
resulting in sharp kinks in the corresponding static structure factors. 
%

Our theoretical results show that the coupling with bosons offers a surprising direct  path to experimentally observe  the elusive Fermi superfluid  using ultracold atoms in smooth one dimensional  flat-bottom traps. The mixture can be brought close to the phase-separation point by tuning one of the three interaction strengths.

Importantly, the combined use of  the full microscopic model together  with accurate  (DMRG) numerics turned out to be essential to unveil the novel effect. 	

While the density modulations in Fig.\ref{fig:fig2_denprof}  vanish for 1D mixtures of infinite size, a weak 
interchain hopping is expected to establish true long range (FFLO) order in higher dimensions. This, in turn, could drive
the formation of bosonic supersolid phases in Bose-Fermi mixtures near phase separation.

\section*{ACKNOWLEDGEMENTS} 
We  acknowledge C. Salomon, F. Chevy, L. Mathey, T. Sowi\' nski and D. Pe\c{c}ak for fruitful discussions. This work was supported by ANR (grant SpifBox).
M.S. acknowledges funding from MULTIPLY fellowships under the Marie Sk\l{}odowska-Curie COFUND Action (grant agreement
No. 713694).

\section*{Appendix A}

\emph{Fourier analysis}. Further insights into the behavior of the mixture can be obtained through a Fourier analysis of the density profiles of bosons and fermions  plotted in Fig.\ref{fig:fig2_denprof} (a)-(b). In order to  avoid boundary effects, we consider only the central region of the chain,
corresponding to $m=44$ sites (approximately one third of the total length).

\begin{figure*}
\centering
\includegraphics[width=0.95\textwidth]{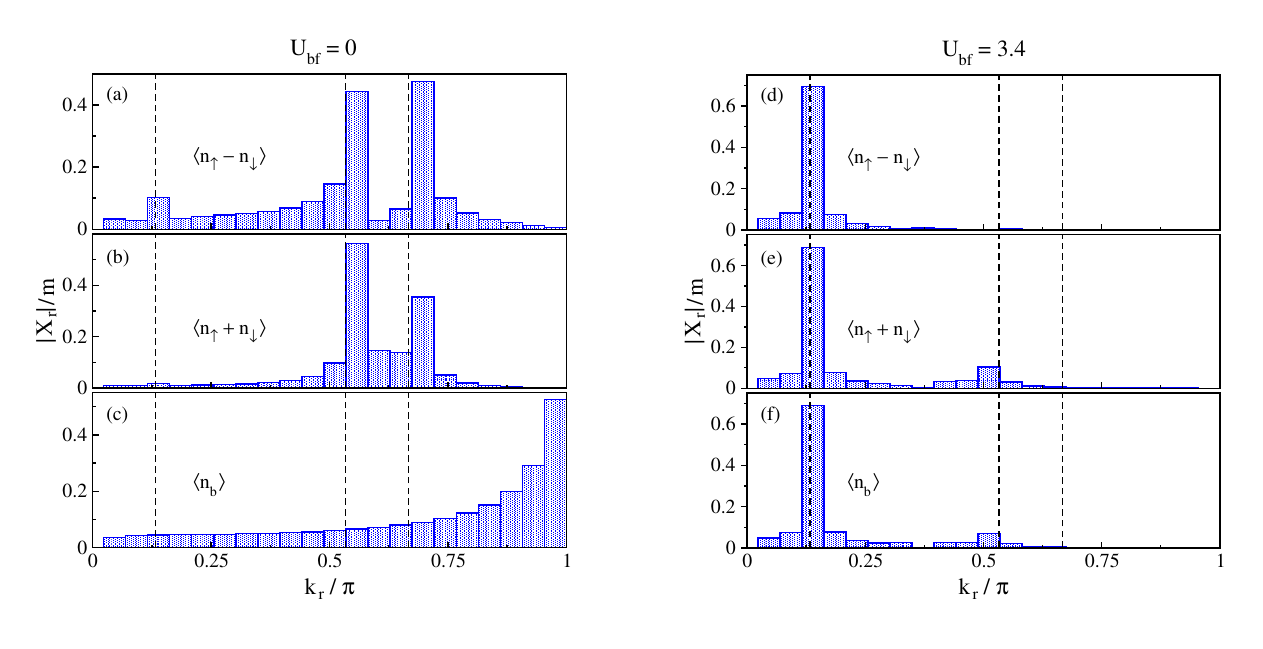}
\caption{(Color online) Fourier analysis of the density profiles of Fig.\ref{fig:fig2_denprof} for $U_{bf}=0$ (left column) and
$U_{bf}=3.4$ (right column) obtained by retaining a window of $m=44$ sites in the middle of the chain: 
the absolute value of the normalized Fourier coefficients $|X_r|/m$ is plotted as a function of the bin wave-vector 
$k_r$ (see text for details). In each column, the three panels from top to bottom display results for the spin density, 
total fermion density, and boson density. The three dashed lines from left to right, refer to 
$2k_{FFLO}, 2k_{F\downarrow}$ and $2k_{F\uparrow}$ momenta, respectively.}
\label{fig:FFT_comb}
\end{figure*}

For each type of  profiles (spin density, fermion total density and boson density) we proceed as follows. Let  $x_i$
denotes the $i$-th element of the truncated data set. 
To avoid uninteresting peaks in the Fourier spectrum we first
normalize our data by setting
 $y_i=(x_i-\mu)/\sigma$, 
where
\begin{equation}
 \mu = {{1}\over{m}}\sum_{i=1}^m x_i,\;\;\;\; \sigma^2 = {\frac{1}{m}\sum_{i=1}^m(x_i-\mu)^2} \nonumber\\
\end{equation}
denote, respectively, the \textit{mean} and the \textit{variance} of the data. 
The corresponding Fourier coefficients are defined as 
\begin{equation}
 X_r = \sum_{n=1}^m y_n e^{-2\pi i r n/m}, \nonumber\\
\end{equation}
where $r=1, 2, ... m$. 
Each integer value $r$  is associated to a bin in Fourier space centered at wave-vector $k_r=2\pi r/m$ and having width  
$2\pi/m$. Since our data are real, it is enough to study the first half of such coefficients, corresponding to $k_r>0$. 

Figs.\ref{fig:FFT_comb}(a)-(c) show the absolute value of the normalized coefficients $X_r/m$, as a function of $k_r$, in the absence of the Bose-Fermi
coupling, $U_{bf}=0$. 
The spin-density (panel a) shows three distinct peaks, corresponding to $2k_{FFLO}=0.133\pi$, $2k_{F\downarrow}=0.533\pi$
and $2k_{F\uparrow}=0.667\pi$. The  $2k_{F\uparrow},2k_{F\downarrow}$ peaks appear also in the spectrum of the total 
fermion density  (panel b), while the FFLO peak is barely visible. The oscillations in the boson density 
profile display instead dominant wave-vector at $2\pi N_b/L=\pi$, as
shown in Fig.\ref{fig:FFT_comb} (panel c). All the above results are consistent with the Luttinger liquid theory.

Next, we repeat the same Fourier analysis for the data in Fig.\ref{fig:fig2_denprof} (b), corresponding to
$U_{bf}=3.4$. 
Close to the phase separation point, all density profiles exhibit a periodic pattern with the same wave-vector,
leading to a  strong peak at $k=2k_{FFLO}$  in the Fourier spectrum, as shown in Figs.\ref{fig:FFT_comb}(d)-(f).
Moreover,  the same figures show that the $2k_{F\downarrow}$  response is significantly reduced, while the 
 $2k_{F\uparrow}$ counterpart  is basically absent. Similar considerations hold for the boson density distribution (panel f), where
  the peak at  $k=\pi$ disappears due to the boson-fermion repulsion.

We emphasize that the outcomes of the above Fourier analysis of the density profiles are fully consistent with the
results for the corresponding static structure factors (SSF) presented in Fig.\ref{fig:fig3_ssf}. In particular the peaks
of the Fourier transform observed in the local densities appear as kinks
in the SSF.

\emph{Finite-size analysis of SSF.}
In Fig.2(c)  we have shown that for $U_{bf}=3.4$ the amplitudes of the FFLO modulations in the local
densities of bosons and fermions scale as the inverse of the system size $L$
and therefore disappear in the termodynamic limit. 
In contrast, the static structure factors of the density operators, displayed in Fig.\ref{fig:fig3_ssf}, remain finite in the thermodynamic limit, as 
they measure spatial correlations in the mixture.

\begin{figure}
	\centering
	\includegraphics[width=0.48\textwidth]{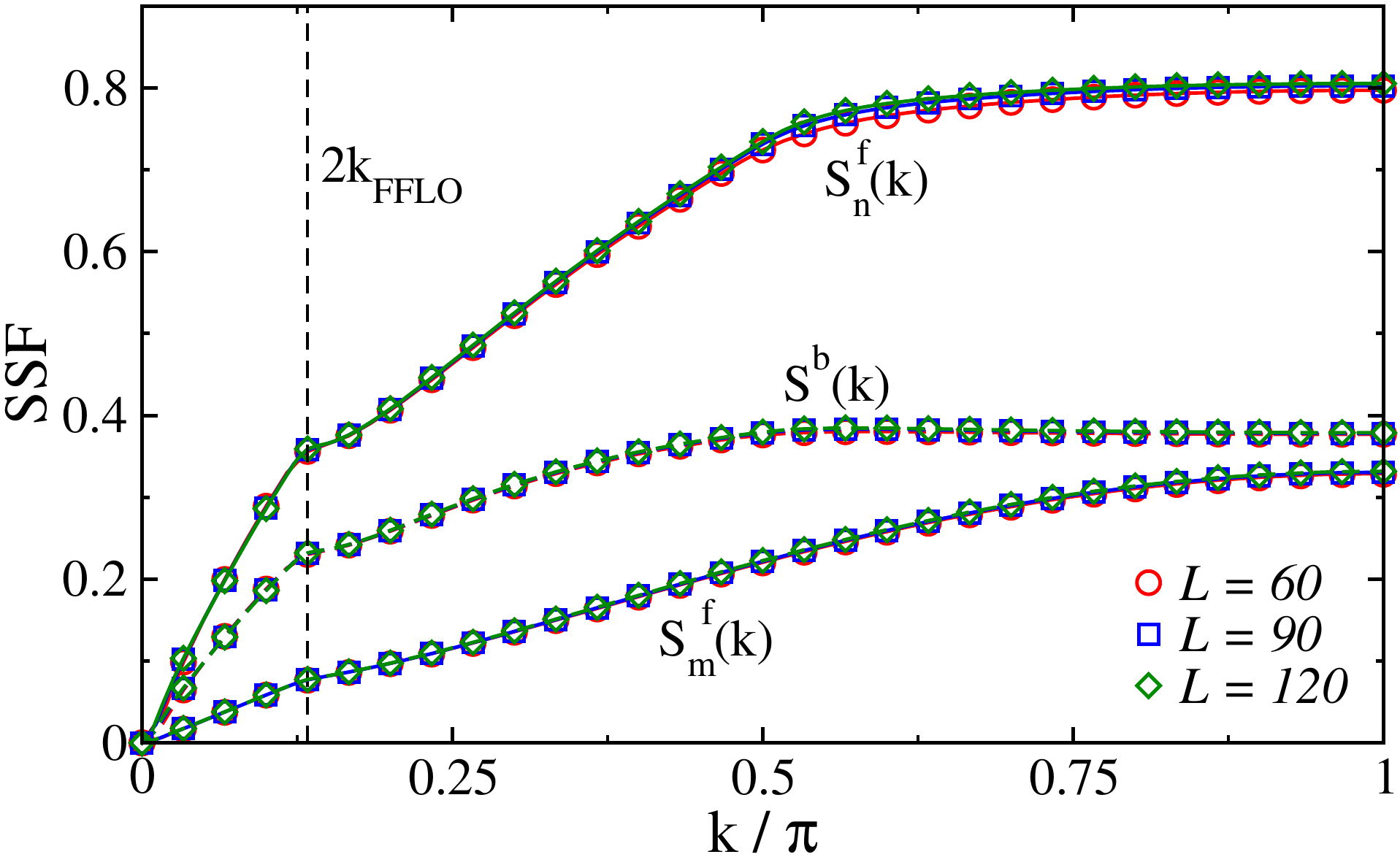}
	\caption{(Color online) Fermion spin density, total density and boson density static structure factors as a function of
		momentum plotted for three different values of the system size, $L=60$ (red circles), $90$ (blue squares) and $120$
		(green diamonds).  All parameters values are the same as in 
		Fig.\ref{fig:fig3_ssf}.}
	\label{fig:scaling_SSF}
\end{figure}

In Fig.\ref{fig:scaling_SSF} we analyze the finite-size effects on the three
SSFs by considering three different lengths of the chain: $L=60$ (red circles), $90$ (blue squares) and $120$
(green diamonds). We see that finite-size effects are indeed negligeable, in particular  the FFLO kinks remain well visibile even for large system sizes.

\section*{Appendix B}
\label{other}

\begin{figure}
	\centering
	\includegraphics[width=0.44\textwidth]{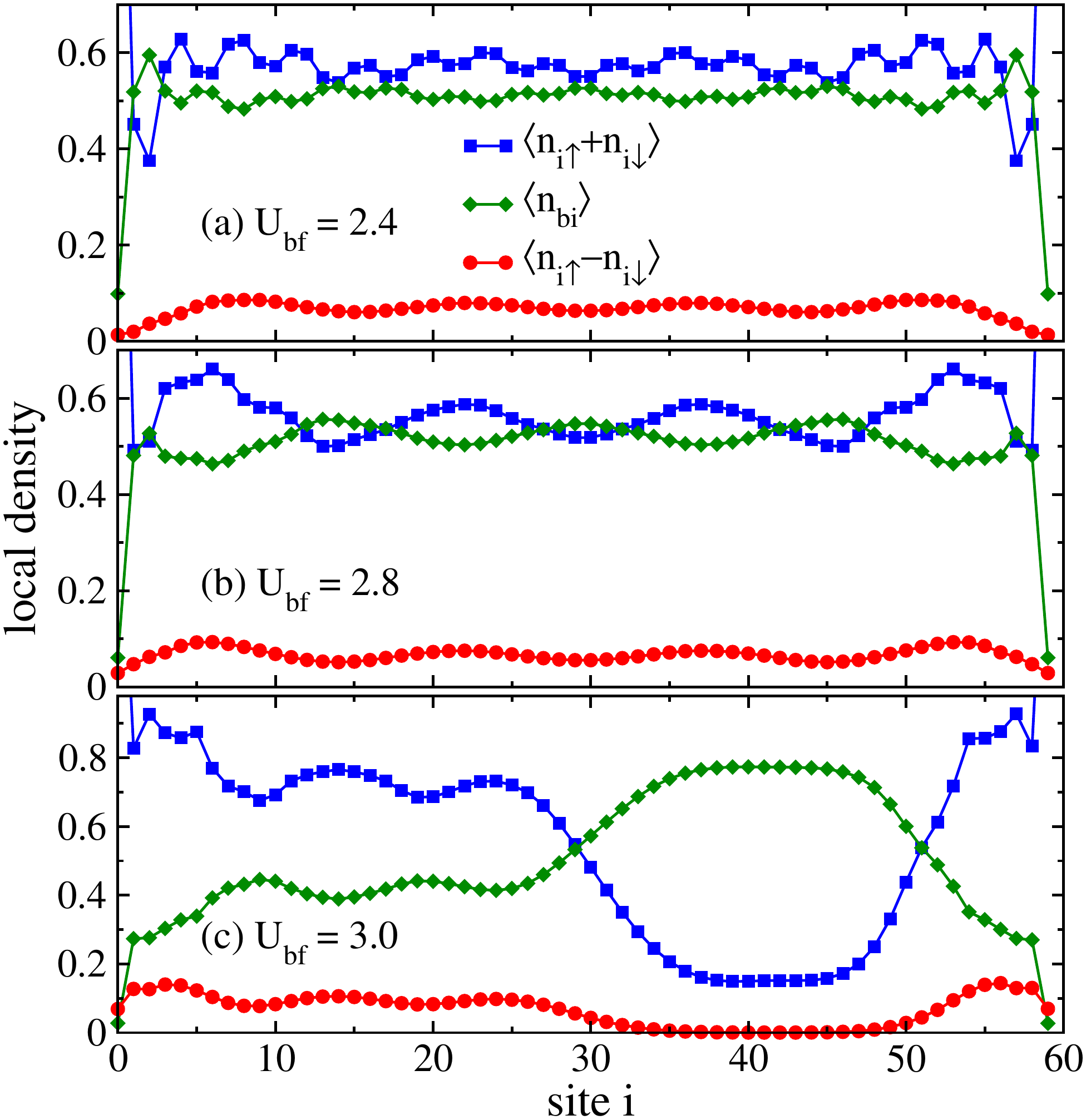}
	\caption{(Color online) Density profiles of bosons and fermions calculated for $U_{bf}=2.4$ (panel a), 
		$2.8$ (panel b) and $3$ (panel c). The Fermi-Fermi interaction strength is fixed at $U_{f}=-5$, while the boson-boson interaction strength 
		and the population numbers are the same as in Fig.\ref{fig:phase_sep1}.}
	\label{fig:phase_sep2}
\end{figure}

\emph{Robustness of the observed effect.} The boson-induced enhancement of the FFLO visibility is a very  general phenomenon, which applies for any set of values of the model parameters such that the Bose-Fermi mixture is close enough to the phase separation limit. In Fig.\ref{fig:phase_sep2} we investigate its appearance  for a mixture with strong fermion-fermion interactions, $U_f=-5$.
 
 As compared to the case $U_{f}=-2$ shown in  Fig.\ref{fig:phase_sep1}, we see
 that phase separation sets in for a weaker boson-fermion repulsion, around $U_{bf}=3$ (panel c). 
For $U_{bf}=2.4$ (panel a), the spin density profile already exhibits the FFLO order, while the other two density profiles  display oscillations with shorter wavelengths. 

For $U_{bf}=2.8$ (panel b),  close to the instability point, the FFLO order is finally imprinted on the density profiles of both species. 
Notice in particular that the sizes of the oscillations amplitudes are fairly similar to those shown in Fig.\ref{fig:phase_sep1} (b). 

Fairly similar results can be obtained by varying the boson-boson interaction. In this case diminishing $U_b$ shifts the phase separation point towards weaker boson-fermion repulsions.

\bibliography{bose-fermiResub.bib}

\end{document}